\documentclass[10pt,twocolumn,letterpaper]{article}

\usepackage[pagenumbers]{iccv} 

\definecolor{iccvblue}{rgb}{0.21,0.49,0.74}
\usepackage[pagebackref,breaklinks,colorlinks,allcolors=iccvblue]{hyperref}
\usepackage{makecell}

\def\confName{ICCV}
\def\confYear{2025}

\begin{document}

\title{FlexPainter: Flexible and Multi-View Consistent Texture Generation}

\author{Dongyu Yan\textsuperscript{\rm 1,$*$}, 
Leyi Wu\textsuperscript{\rm 1,$*$}, 
Jiantao Lin\textsuperscript{\rm 1}, 
Luozhou Wang\textsuperscript{\rm 1}, 
Tianshuo Xu\textsuperscript{\rm 1}, 
Zhifei Chen\textsuperscript{\rm 1}, \\
Zhen Yang\textsuperscript{\rm 1}, 
Lie Xu\textsuperscript{\rm 2}, 
Shunsi Zhang\textsuperscript{\rm 2},
Yingcong Chen\textsuperscript{\rm 1,3,$\dag$}\\
\textsuperscript{\rm 1}HKUST(GZ), \textsuperscript{\rm 2}Quwan, \textsuperscript{\rm 3}HKUST\\
{\tt\small \{starydyxyz@gmail.com; lwu398@connect.hkust-gz.edu.cn; yingcong.ian.chen@gmail.com\}
}
}

\twocolumn[{
\renewcommand\twocolumn[1][]{#1}
\maketitle
\begin{center}
\vspace{-8mm}
    \captionsetup{type=figure}
    \includegraphics[width=0.9\textwidth]{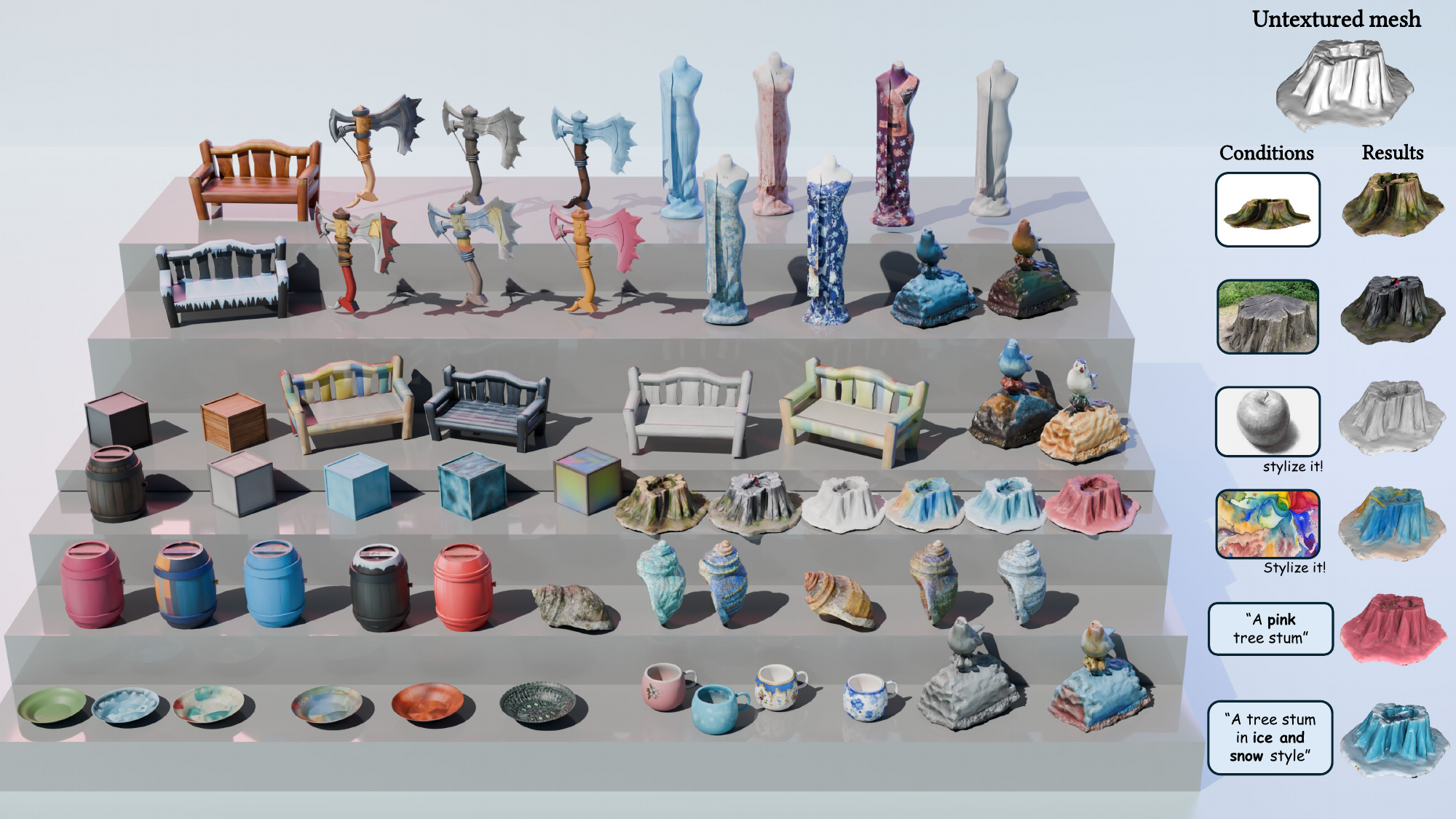}
\vspace{-2mm}
    \captionof{figure}{
        \textbf{FlexPainter} generates diverse, high-quality textures based on various flexible user prompts.
    }
\end{center}
}]

\maketitle
\renewcommand{\thefootnote}{}
\footnote{$*$ Both authors contributed equally to this research.}
\footnote{$\dag$ Corresponding Author.}

\begin{abstract}
Texture map production is an important part of 3D modeling and determines the rendering quality.
Recently, diffusion-based methods have opened a new way for texture generation.
However, restricted control flexibility and limited prompt modalities may prevent creators from producing desired results.
Furthermore, inconsistencies between generated multi-view images often lead to poor texture generation quality.
To address these issues, we introduce \textbf{FlexPainter}, a novel texture generation pipeline that enables flexible multi-modal conditional guidance and achieves highly consistent texture generation.
A shared conditional embedding space is constructed to perform flexible aggregation between different input modalities.
Utilizing such embedding space, we present an image-based CFG method to decompose structural and style information, achieving reference image-based stylization.
Leveraging the 3D knowledge within the image diffusion prior, we first generate multi-view images simultaneously using a grid representation to enhance global understanding.
Meanwhile, we propose a view synchronization and adaptive weighting module during diffusion sampling to further ensure local consistency.
Finally, a 3D-aware texture completion model combined with a texture enhancement model is used to generate seamless, high-resolution texture maps.
Comprehensive experiments demonstrate that our framework significantly outperforms state-of-the-art methods in both flexibility and generation quality.
\end{abstract}    
\section{Introduction}

Texturing is a crucial step in 3D modeling and an important aspect of computer graphics and vision.
It has significant applications in games, film, VR/AR, and animation.
With the development of diffusion models\cite{saharia2022photorealistic, ramesh2022hierarchical, rombach2022high, podell2023sdxl, esser2024scaling, flux2023}, works utilizing image generation models for texture generation have demonstrated strong potential \cite{richardson2023texture, chen2023text2tex, wu2024texro, xiang2024make, gao2024genesistex, huo2025texgen}.
However, extending these methods from general image generation to texture creation remains challenging.

A key challenge in multi-view-based texture generation is precisely controlling the generated results according to user expectations.
Most of the existing texture generation methods, like TEXTure \cite{richardson2023texture}, Text2Tex \cite{chen2023text2tex}, and MVPaint \cite{cheng2024mvpaint} use text as the condition.
Other works like Paint3D \cite{zeng2024paint3d} and FlexiTex \cite{jiang2024flexitex} utilize IP-Adapter \cite{ye2023ip} to allow image-conditioned generation.
Although these methods are able to generate high-quality texture maps and can accept multiple modalities, they are restricted to conditioning on only one modality at a time, which may cause ambiguity.
For example, text descriptions may fail to convey detailed color, style, or material information accurately.
Meanwhile, finding an image condition that perfectly meets the user's demand is difficult; mixed information within it may lead to unclear semantic intent and missing contextual details.
Differently, StyleTex \cite{xie2024styletex} uses a reference style image to generate stylized textures.
However, using an additional GPT model for content-style separation increases the system's complexity, and the style feature obtained may preserve unwanted structural information of the input image.

Another important challenge of texture generation is ensuring inter-view consistency, as global and local discrepancies can significantly degrade the plausibility and detail quality of the generated texture.
Previous works \cite{richardson2023texture, chen2023text2tex} usually generate one viewpoint at a time sequentially to paint the mesh.
These generated views are independent, which may easily lead to inconsistencies.
To overcome such problems, methods like SyncMVD \cite{liu2024text} and TexFusion \cite{cao2023texfusion} merge different views in UV space during diffusion sampling and achieve view-synchronized generation.
Although these methods can produce locally consistent images, there is still no information exchange between non-overlapping views, which causes the Janus problem.
To gain holistic information across views, Meta TextureGen \cite{bensadoun2024meta} and MVPaint \cite{cheng2024mvpaint} 
leverages a multi-view image grid as the generation target.
However, after reprojecting into UV space, misalignments in the details may still occur, resulting in ghosting artifacts.
Moreover, both types of the above methods employ a heuristic weight map to merge partial textures reprojected from different views.
Such a simple weighting function can cause low robustness and adaptability to different inputs.

To address the aforementioned challenges, we introduce FlexPainter, a novel texture-generation pipeline with flexible input and consistent generation.
Leveraging the shared, linear-structured conditional embedding space of T5 \cite{raffel2020exploring}, we achieve the complementary integration of high-level semantics from texts and low-level details from images.
Such multi-granularity control helps generate textures that better meet designers' needs in real-world production, reducing the time spent on repeated adjustments.
With such embedding space, we further inject image embedding into the classifier-free guidance (CFG) \cite{ho2022classifier} module.
Such image-based CFG not only improves generation quality but also provides flexible control.
It can eliminate the structure and content information in the image prompt by feeding grayscaled versions as negative prompts, yielding aesthetic texture stylization. 

For inconsistency, our method handles it from both global and local perspectives.
We apply the multi-view image grid representation as the generation target to gain the model with global understanding.
This way, attention between views can ensure a holistic understanding of the object, thereby maintaining global coherence.
As for the local alignment, we apply view synchronization during diffusion sampling.
At each denoising step, we reproject views back to UV space, aggregate them into a unified UV map, and then rasterize it into the image grid representation for the next step of generation.
We use an additional adaptive WeighterNet to aggregate the partial UV maps.
Compared with the traditional heuristic weighting function used in \cite{bensadoun2024meta, cheng2024mvpaint, zhang2024texpainter}, our method can handle various generated inputs at different timesteps and output robust results.
Finally, a diffusion-based texture completion module and a texture enhancement network are applied to obtain the full texture map with high-resolution details.

In summary, our contributions are as follows:
\textbf{(1)} We propose FlexPainter, a carefully designed pipeline for generating high-quality and consistent textures, offering intuitive granularity control.
\textbf{(2)} We conduct linear operations and image-based CFG in the conditional embedding space, achieving flexible generation and coherent stylization based on different user demands and prompt modalities.
\textbf{(3)} We realize global and local consistent texture generation through a unified image grid representation combined with a view synchronization and weighting module.
\section{Related Work}

\subsection{Multi-View Generation}

Multi-view generation tasks aim to produce multiple perspectives of an object with the reference image or text.
Zero-1-to-3 \cite{liu2023zero} and Consistent-1-to-3 \cite{ye2024consistent} generate novel views using a viewpoint-conditioned manner.
Zero123++ \cite{shi2023zero123++}, MVDream \cite{shi2023mvdream}, and other works \cite{li2023instant, wang2023imagedream, long2024wonder3d, wu2024unique3d} manage to generate multi-view grid images simultaneously for better consistency.
Some works \cite{voleti2025sv3d, xu2023dmv3d, melas20243d, liu2023syncdreamer, weng2023consistent123, huang2024mv, xu2024flexgen} incorporate various modules, including video diffusion priors, iterative view synchronization with 3D proxy, and flexible prompt-based input to accomplish the task.
Unlike the above methods, which mainly focus on subsequent 3D generation, multi-view generation for texturing incorporates additional geometry constraints.
It requires multi-view information to be projected and merged into UV space to obtain the final texture map, imposing stricter requirements on multi-view consistency.

\subsection{Texture Generation}

Texture generation aims to produce UV texture maps of the target mesh with the given prompts as conditions.
Traditional learning-free methods \cite{johannes2007solid, lefebvre2006appearance, turk2001texture, wei2009state} rely on manual or algorithm-specific procedures to generate textures.
Approaches like \cite{oechsle2019texture, siddiqui2022texturify, bokhovkin2023mesh2tex, cheng2023tuvf} performs texture generation based on GAN structure \cite{goodfellow2020generative, karras2020analyzing}.
With the rapid development of diffusion models, their adoption in texture generation has been explored.
Approaches like \cite{poole2022dreamfusion, lin2023magic3d, chen2023fantasia3d, metzer2023latent, yeh2024texturedreamer} optimize texture fields based on Score Distillation Sampling (SDS).
Other works \cite{richardson2023texture, chen2023text2tex, perla2024easitex, wu2024texro, xiang2024make, zhao2025hunyuan3d} use pre-trained diffusion models to generate in the image domain and apply predefined rules to fuse images from different views into the UV domain.

\subsubsection{Condition flexibility}

Many texture generation methods attempt to allow multi-modality input to better guide texture generation.
TEXTure \cite{richardson2023texture} and TextureDreamer \cite{yeh2024texturedreamer} achieve concept transfer using textual inversion \cite{gal2022image} and
DreamBooth \cite{ruiz2023dreambooth} techniques.
In addition to basic text input, FlexiTex \cite{jiang2024flexitex}, Paint3D \cite{zeng2024paint3d} and MVEdit \cite{chen2024generic} allow for image conditioning by adopting IP-Adapter \cite{ye2023ip} module.
However, these works still use fixed inputs and cannot integrate multiple conditional controls flexibly.
Differently, StyleTex \cite{xie2024styletex} proposes to operate embeddings in the CLIP \cite{radford2021learning} space to achieve stylized texture generation.
While aiming to disentangle content and style, the additional language model introduces overhead and may undesirably preserve structural information from the input image.

\subsubsection{UV space consistency}

Direct back-projection of the diffusion-generated multi-view images may cause misalignments in UV space.
To address the UV inconsistencies from different generated views, methods like TexFusion \cite{cao2023texfusion}, SyncMVD \cite{liu2024text}, and others \cite{jiang2024flexitex, zhang2024texpainter, gao2024genesistex, wu2024texro, liu2024vcd, huo2025texgen} share noisy content across views via UV projection during denoising loop.
However, these methods struggle to maintain global consistency due to insufficient holistic information, often producing artifacts and Janus Problem.
Other works \cite{bensadoun2024meta, cheng2024mvpaint, deng2025flashtex} leverage image grid representation for simultaneous view generation.
Although attention between views can enhance global consistency, it may lead to local misalignment, affecting aggregated UV results.
Point-UV \cite{yu2023texture}, and TEXGen \cite{yu2024texgen} take a different approach by directly generating texture maps using point and UV hybrid structures.
However, they are trained from scratch, relying on limited 3D datasets, leading to limited generalizability.

\section{Preliminaries}

\subsection{Latent Flow Matching Method}

The flow matching method \cite{lipman2022flow} is a generative method that transforms a noise distribution into a target distribution.
It is modeled by velocity prediction that guides sampling along a smooth trajectory in data space.
Specifically, the most commonly used rectified flow defines a straight line as the path:
\begin{equation}
    x_t = (1-t)x_0 + t\epsilon, \quad \frac{dx}{dt} = v_t(x_t, \Theta),
\end{equation}
where $x_t$ is a sample point on the flow trajectory in time step $t$, $x_0$, $\epsilon$ are the destinations from two distributions, and $v_t$ represents the velocity field modeled by a neural network with parameters $\Theta$.

Flow matching-based image generation is typically extended to the latent space to improve scalability and efficiency.
To this end, a variational autoencoder (VAE) \cite{kingma2013auto} is used for the mapping between image space and latent space:
\begin{equation}
    x_0 = \mathcal{E}(I), \quad I = \mathcal{D}(x_0),
\end{equation}
where $\mathcal{E}$ and $\mathcal{D}$ denotes the encoder and decoder of the VAE, and $I$ is a sample image in the original space.

\subsection{Texture Rasterization and ReProjection}

Texture rasterization and reprojection are fundamental techniques in computer graphics, enabling mapping between rendered images and UV textures.
Rasterization involves projecting a 3D model onto a 2D image plane, determining visibility, and assigning texture values from UV space to image pixels.
In contrast, reprojection involves back-projecting images from a certain view onto the 3D object, which reconstructs textures based on known depth and camera information.
With camera $c$ and texture map $T$, they can be noted as:
\begin{equation}
    I = \mathcal{R}^(T,c), \quad T = \mathcal{R}^{-1}(I,c).
\end{equation}

\section{Method}

\begin{figure}[t]
    \centering
    \includegraphics[width=0.5\textwidth]{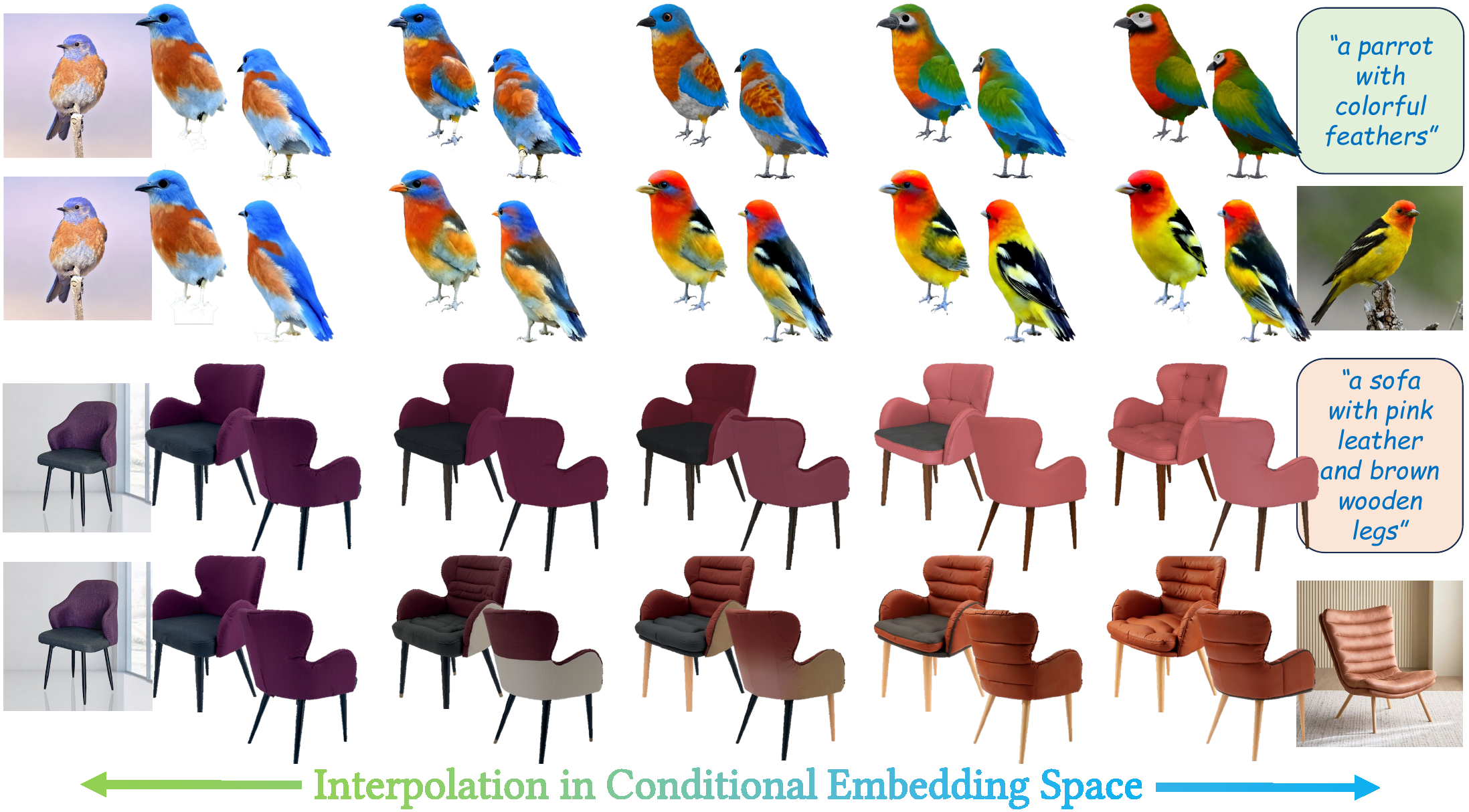}
    \caption{
        Example of modality aggregation and semantic manipulation by controlling the image embedding strength $\alpha$.
        Each case's first and second rows show text-to-image and image-to-image interpolation, respectively.
    }
    \label{fig:inter}
\end{figure}

\begin{figure*}[t]
    \centering
    \includegraphics[width=1\textwidth]{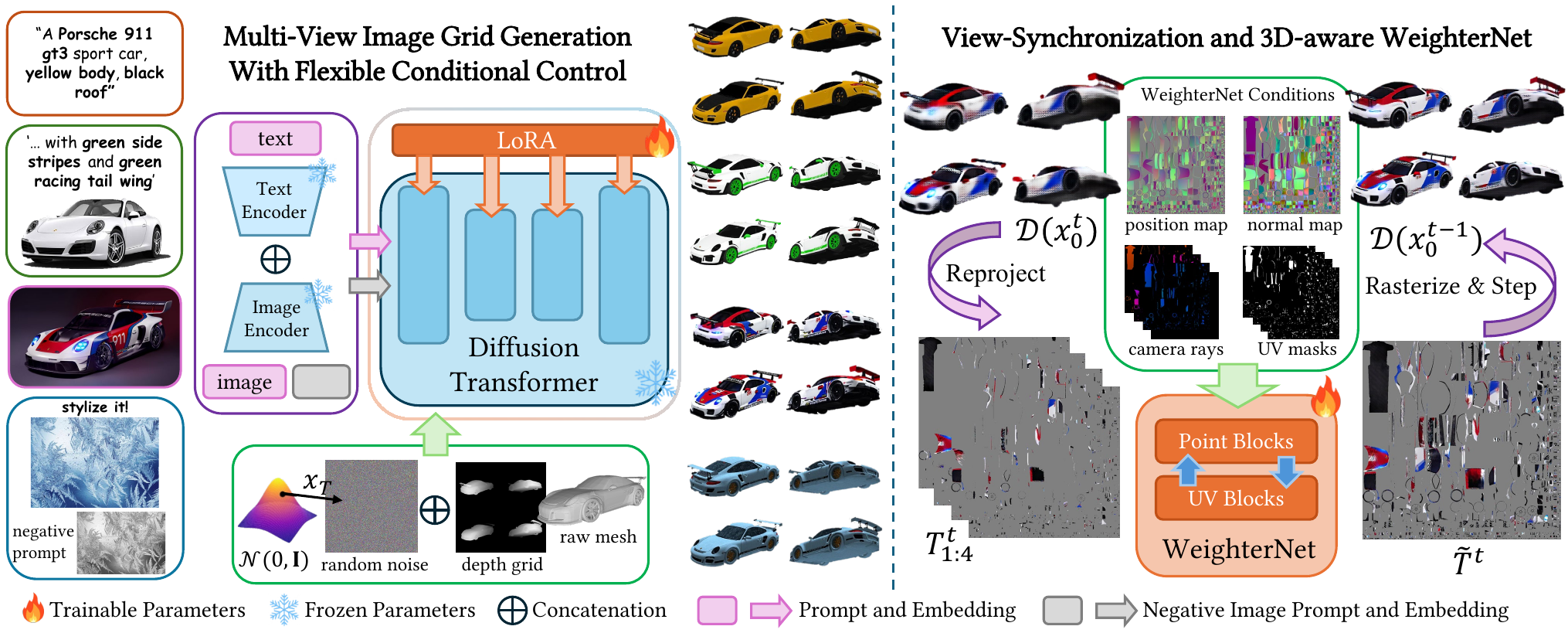}
    \caption{
        Pipeline of our method.
        We first generate multi-view images using the conditional input from the user.
        The top two cases show the generation of using text-only or image-only conditions.
        Leveraging the linear operation in Equ.(\ref{equ:lo}), we can also perform text-guided image refinement (shown in green) and stylization using reference image (shown in blue).
        The right side shows our view-synchronization and weighting module.
        Consistent multi-view images can be generated by reprojection and weighting during each sampling step.
    }
    \label{fig:pipeline}
\end{figure*}

Given a 3D object mesh, we aim to generate high-fidelity textures that reflect users' creative intent through diverse prompts.
We implement a prompt aggregation method on a multi-modal embedding space to achieve flexible conditioning.
We also achieve aesthetic texture stylization with an image-based CFG method (Sec. \ref{sec:cond}).
As for the generation pipeline, we use a multi-view image grid representation to strengthen global understanding.
We propose a reprojection-based view synchronization and weighting module during diffusion sampling for better view consistency (Sec. \ref{sec:pipe}).
The full pipeline of our method is shown in Fig. \ref{fig:pipeline}.

\subsection{Flexible Conditioning}
\label{sec:cond}

\subsubsection{Aggregation of multi-modal prompts}

Accommodating multi-modal inputs to achieve desired results flexibly is a challenging topic.
To enable text, images, and their interactions as conditions, we use T5 \cite{raffel2020exploring} as a shared embedding space to align the information from each modality.
Text and image inputs are encoded by separate encoders into this space, and their embeddings are concatenated to form a unified condition that guides the generation process via cross-attention.
Thanks to the linear properties of the T5 embedding space, we can manipulate the strength and direction of the image embeddings to achieve flexible control. The final embedding $\mathbf{v}$ is obtained as:
\begin{equation}
    \mathbf{v} = \mathcal{T}(w_{cond}) \bigoplus \sum_{i=1}^{n} \alpha_i \mathcal{I}(I^i_{cond}),
\label{equ:lo}
\end{equation}
where $\mathcal{T}$ and $\mathcal{I}$ represents the text and image encoders respectively, $w_{cond}$ and $I^i_{cond}$ are the text and $i$-th image condition.
Such manipulation in embedding space results in meaningful generations, as illustrated in Fig. \ref{fig:inter}.
Even with single-modal training, the linear structure and combination property can be well preserved, allowing advanced tasks such as text-assisted image prompt refinement and generation using reference-style images to be effectively implemented without special design.

\subsubsection{Image-based CFG}

Traditional CFG is usually carried out using blank text, which lacks strong and precise control.
Utilizing the above multi-modal embedding space, we can inject image information into the diffusion network to enhance its generative capabilities.
Furthermore, we incorporate image embeddings into the negative prompts used in the CFG framework \cite{ho2022classifier} to improve generation control.
Employing image embeddings yields better conditional control and higher-quality generation results than textual-only embeddings as negative prompts.
Moreover, we can utilize images containing undesired structures or specific colors as negative prompts to explicitly suppress their presence in the generated outputs.
Building on this property, we further achieve reference image-conditioned texture stylization.
Due to the entanglement between content and style, directly using the original reference image as a prompt may introduce unintended structural artifacts into the generated texture.
To address this, we employ a grayscaled version of the reference image as a negative prompt to eliminate structural information while preserving stylistic features. This disentanglement of content and style enables more robust and flexible texture stylization, allowing the model to focus on transferring color and tone without copying shape or layout from the reference.

\subsection{Generation Pipeline}
\label{sec:pipe}

\subsubsection{Representation of the multi-view images}
We first perform generation in the image space by leveraging the strong priors in current state-of-the-art 2D diffusion models.
However, generating a single image per sample can cause global inconsistencies due to the lack of communication between perspectives.
We adopt image grid representation for synchronized multi-view generation in a single sample to facilitate information exchange between different views and maximize the use of 3D prior information from existing image diffusion models.
Specifically, we arrange four evenly spaced surround views of the object in a two-by-two grid as our generation target.
For geometry alignment, depth maps are rendered and concatenated with the noisy color images to form the input to the diffusion network.
In particular. we use LoRA \cite{hu2021lora} to retarget the pre-trained model to generate pure background and coherent multi-view grid images.
The image grid representation and simultaneous generation strategy allow attention between views, which can better enhance global understanding and consistency.
The 3D knowledge within the model's large amount of training data makes it much easier to adapt to the new task.

\subsubsection{Reprojection-based view synchronization}

The use of image grid representation allows us to generate globally consistent multi-views.
However, local misalignments can still occur when the images are reprojected into UV space, affecting the overall texture quality.
To address this problem, we propose a reprojection-based view-synchronization module to align different views in UV space at each sampling step.
Specifically, after obtaining the predicted flow $v_t$ from the noisy latent $x_t$ at denoising step $t$, we first use it to predict the fully denoised latent $x^t_0$:
\begin{equation}
\label{equ:flow}
    x^t_0 = x_t - tv_t(x_t).
\end{equation}
Then, the latent $x^t_0$ is decoded, split into independent images, and reprojected into UV space, producing four partial texture maps $T^{t}_{1:4}$:
\begin{equation}
\label{equ:project}
    \begin{split}
        I^{t}_{1:4} = \mathcal{S}&(\mathcal{D}(x^t_0)), \\
        T^{t}_i = \mathcal{R}^{-1}(I^{t}_i, c_i)&, \quad i = 1,2,3,4,
    \end{split}
\end{equation}
where $\mathcal{S}$ denotes splitting operation that turns grid image into individuals images $I^t_{1:4}$, and $c_i$ represents the $i$-th camera.
Next, a weighting function $\mathcal{W}$ is applied on the partial texture maps $T^{t}_{1:4}$ to generate the fused texture map $\tilde{T}^{t}$:
\begin{equation}
    \tilde{T}^{t} = \mathcal{W}(T^{t}_{1:4}).
\end{equation}
From $\tilde{T}^{t}$, the view-synchronized latent $\tilde{x}^t_0$ can be obtained through operations inverse to the above process:
\begin{equation}
    \tilde{I}^{t}_{1:4} = \mathcal{R}(\tilde{T}^{t}, c_{1:4}), \quad \tilde{x}^t_0 = \mathcal{E}(\mathcal{F}(\tilde{I}^{t}_{1:4})).
\end{equation}
Finally, a view-synchronized flow prediction $\tilde{v}_t$ is computed to replace the original $v_t$, and the subsequent sampling steps continues:
\begin{equation}
    \tilde{v}_t = \frac{x_t - \tilde{x}^t_0}{t}.
\end{equation}
This reprojection-based view synchronization ensures that, at each diffusion step, the multi-view images to be denoised are derived from the same UV map.
This way, the consistency between views can be further maintained.

\begin{figure*}[t]
    \centering
    \includegraphics[width=1\textwidth]{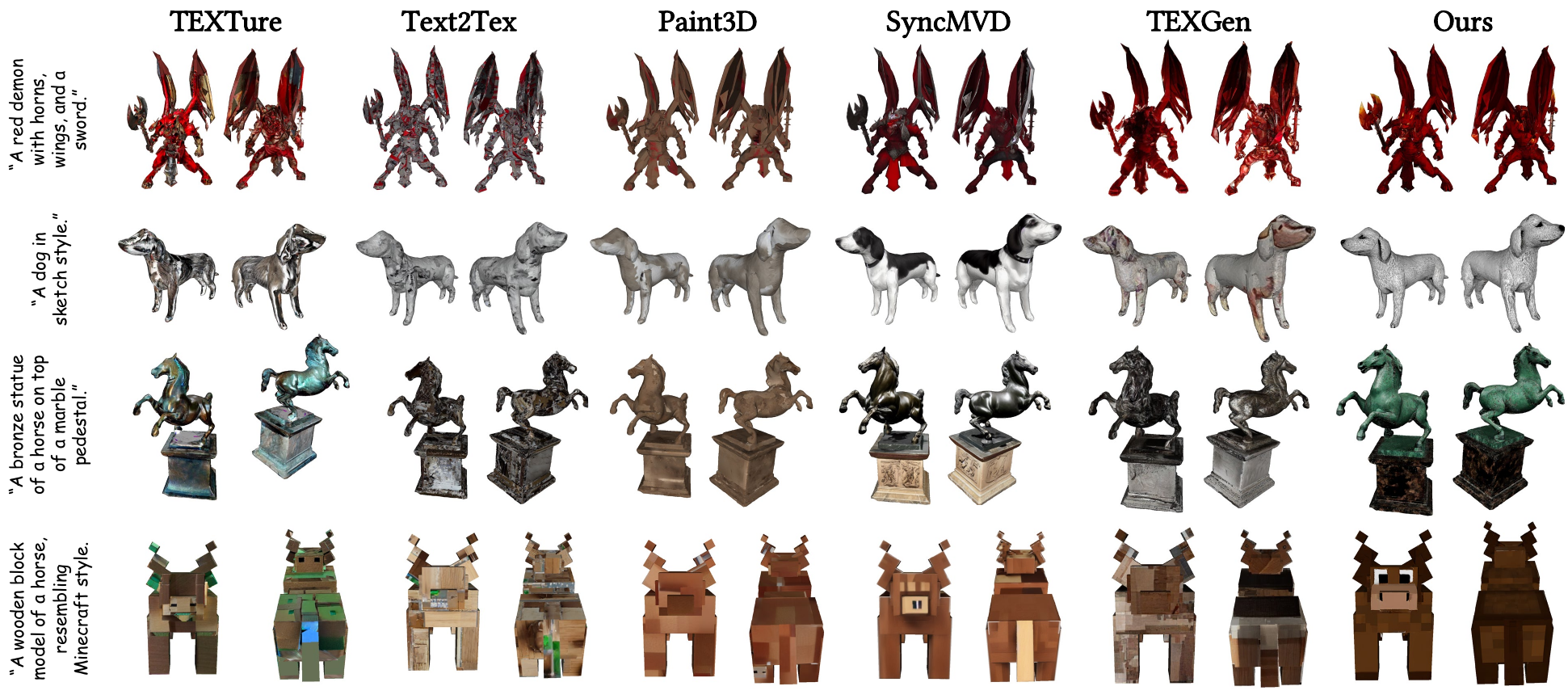}
    \caption{
        Qualitative results on text-to-texture generation.
    }
    \label{fig:text}
\end{figure*}

\subsubsection{Adaptive Weighting}

Although multi-view consistency can be strengthen using view synchronization, the generation quality can be greatly affected by the choice of the weighting function $\mathcal{W}$.
Traditionally, the weights assigned to partial textures from different views are determined heuristically, using the cosine of the angle between the camera rays and the rendered normals of the mesh as the weighting factor.
While such a weighting function provides a simple way to detect observation quality, it can not dynamically adjust to different input variations.
It may introduce poorly generated regions into the final result.
Therefore, we train a WeighterNet as the weighting function $\mathcal{W}$, allowing it to discern generation quality and achieve robust weighting results.
Due to the discontinuities in UV space, incorporating 3D structures to enhance the network's spatial awareness is beneficial.
Thus, we combine the UV blocks and point blocks by the mapping between 2D and 3D space.
Each UV block encodes global context in the 2D UV space, after which features are projected onto 3D points via known UV mappings. A point block then aggregates geometric information across neighboring points. The updated features are reprojected back to the UV map for further refinement. These processes are repeated to generate the final weights for multi-view aggregation.
The WeighterNet also considers the input variation of different time steps by taking an additional $t$ as input.
The final format can be represented as:
\begin{equation}
    \tilde{T^t} = \mathcal{W}(T^t_{1:4}, R_{1:4}, N_{1:4}, P, t, \Theta),
\end{equation}
where $P$ represents the position map in the UV space.
We use ground truth textures $\bar{T}$ and its simulated denoised texture $\tilde{T}^t$ at time step $t$ as a training pair.
For supervision, perceptual loss and cycle loss are used:
\begin{equation}
    \begin{split}
        L_{pec} =& e^{-\alpha t} \sum_{i=1}^{4} \text{vgg}(\mathcal{R}(\bar{T}, c_i)) - \text{vgg}(\mathcal{R}(\mathcal{W}(T^t_{1:4}, \Theta), c_i)), \\
        L_{cyc} =& \sum_{i=1}^{4} \mathcal{R}(T_i^t, c_i) - \mathcal{R}(\mathcal{W}(T^t_{1:4}, \Theta), c_i),
    \end{split}
\end{equation}
where $e^{-\alpha t}$ balances the penalties for different noise levels.
As a result, our adaptive weighting function can provide adaptable view aggregation in a training-based manner.

Although the multi-view diffusion model has attention between each view, it lacks strict consistency restrictions. The mechanism of our WeighterNet is to enhance such restriction in an adaptive way, treating multi-view images as a whole. The network are meant to be trained along with the diffusion model. However, due to the limitation of the memory, we train the WeighterNet alone using the one-step denoised $x_0$ images generated by the trained multi-view diffusion model.

\section{Experiment}

We conduct qualitative and quantitative experiments to verify the superiority of our method.
Ablation studies are also performed to prove the effectiveness of our proposed modules.
Furthermore, we demonstrate examples of our method in practical applications to illustrate its real-world significance.

\subsection{Implementation Details}

We choose FLUX.1-dev \cite{flux2023} as our base generation model and utilize LoRA \cite{hu2021lora} to retarget it for multi-view image grid generation.
We use T5 \cite{raffel2020exploring} text encoder and Redux \cite{flux2023} image encoder to align multi-modal information in a shared embedding space.
For the WeighterNet, we use UNet structure for 2D and  point attention for 3D to achieve UV and spatial perception.
To avoid severe detail loss, we only apply adaptive weighting for the first half of the steps.
For texture completion and enhancement, we leverage the inpainting capabilities of TEXGen \cite{yu2024texgen} and finetune it on our synthesized partial texture data.
We then deploy Real-ESRGAN \cite{wang2021real} finetuned on the UV map dataset to achieve texture refinement and super-resolution.

\subsubsection{Data preparation}

We select 100K 3D objects from the Objaverse dataset \cite{deitke2023objaverse} as our training set.
During training, we randomly select rendered images and GPT-4V labeled caption as prompt.
Four surrounding, relatively fixed depth and albedo images are also rendered as training data.
For WeighterNet training, we simulate denoised texture maps $\tilde{T}^t$ at time step $t$ by first adding noise on the ground-truth rendered views to get $x_t$.
Then, we can use Equ. (\ref{equ:flow}) and (\ref{equ:project}) to get the simulate $\tilde{T}^t$ as training input.

\subsubsection{Baselines and metrics}

We choose TEXTure \cite{richardson2023texture}, Text2Tex \cite{chen2023text2tex}, SyncMVD \cite{liu2024text}, Paint3D \cite{zeng2024paint3d} and TEXGen \cite{yu2024texgen} as our comparison baseline In particular, we use a depth-controlled Flux \cite{flux2023} model to generate the mesh-aligned front view required by TEXGen to ensure fairness.
We randomly select 100 unseen 3D objects from Google Scanned Objects (GSO) \cite{downs2022google} as the evaluation dataset to better demonstrate our generalization capabilities.
Common metrics, including FID \cite{heusel2017gans} and KID \cite{binkowski2018demystifying}, are used for quantitative comparison.
To further prove our effectiveness, we conduct a user study by asking users to select the texture generation result with the highest quality that best matches the text or image prompt.
The results are represented as the proportion of user preference.

\begin{figure}[t]
    \centering
    \includegraphics[width=0.5\textwidth]{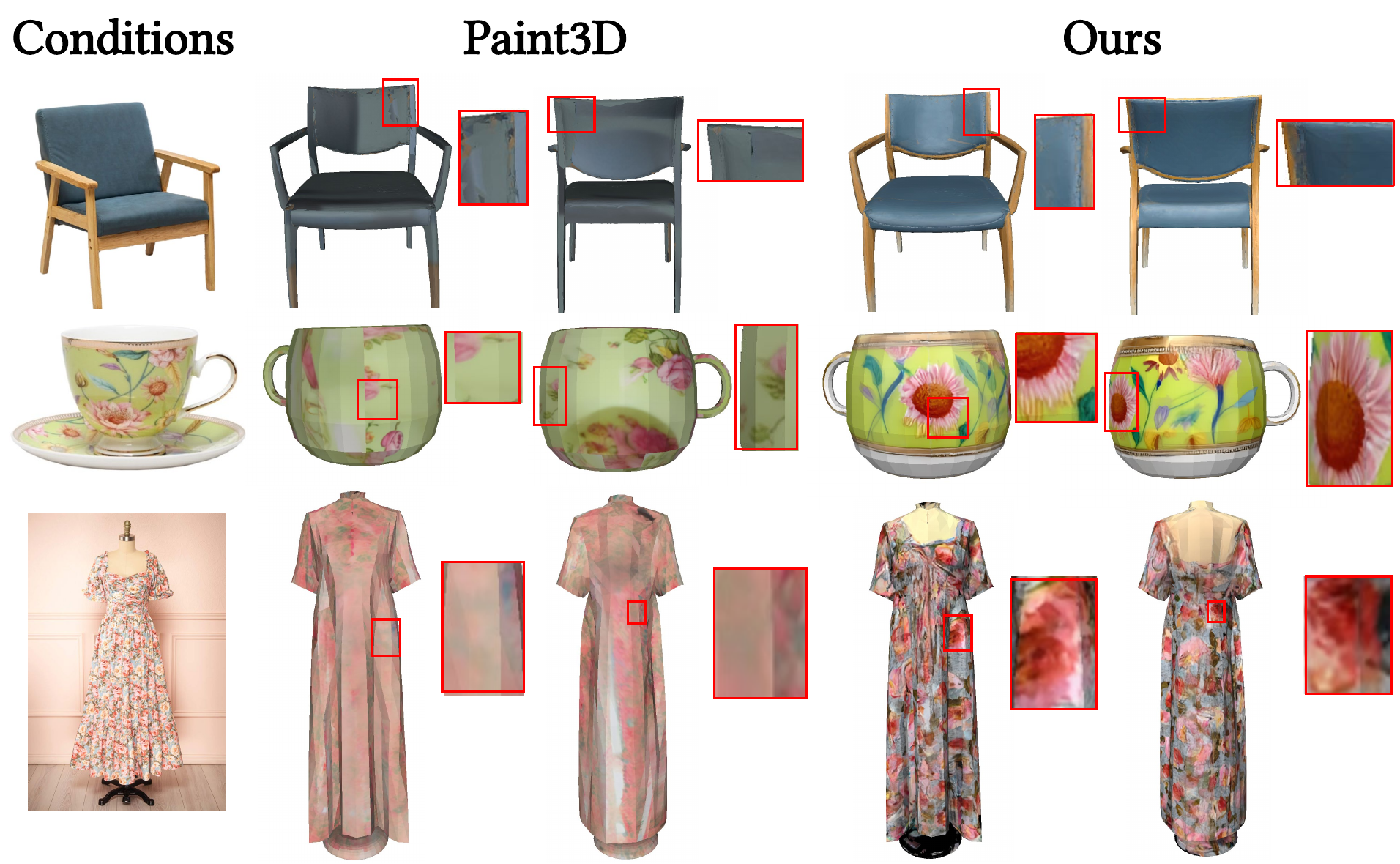}
    \caption{
        Qualitative results on image-to-texture generation.
    }
    \label{fig:image}
\end{figure}

\subsection{Comparisons on Text-to-Texture}

\begin{table}[t]
\caption{
    Quantitative evaluation result of text-to-texture generation.
    The best results are in bold, while the second-best ones are underlined.
    KID metric is multiplied by ${10}^{-4}$ and the unit of runtime is seconds.
}
\centering
\begin{tabular}{ccccc}
\hline
& FID $\downarrow$ & KID $\downarrow$ & User Preference & Runtime \\ 
\hline
TEXTure & 78.269 & 107.062 & 4.4\% & 251 \\ 
Text2Tex & 89.008 & 94.423 & 11.3\% & 996 \\ 
SyncMVD & 73.385 & \textbf{41.737} & \underline{22.5\%} & 113 \\ 
Paint3D & 77.801 & 123.605 & 17.8\% & 204 \\ 
TEXGen & \underline{72.900} & 61.729 & 15.7\% & 52 \\ 
Ours & \textbf{71.621} & \underline{58.465} & \textbf{28.3\%} & 153 \\ 
\hline
\end{tabular}
\label{tab:text}
\end{table}

\begin{table}[t]
\centering
\caption{
    Quantitative evaluation result of image-to-texture generation.
    The best results are indicated in bold.
    KID metric is multiplied by ${10}^{-4}$ and the unit of runtime is seconds.
}
\begin{tabular}{ccccc}
\hline
& FID $\downarrow$ & KID $\downarrow$ & User Preference & Runtime \\ 
\hline
Paint3D & 83.977 & 267.132 & 28.6\% & 317 \\ 
Ours & \textbf{59.492} & \textbf{62.089} & \textbf{71.4\%} & 157 \\ 
\hline
\end{tabular}
\label{tab:image}
\end{table}

We conduct experiments on text-to-texture generation.
We only use text as the condition for this task and set all $\alpha_i$ to 0.
The results are shown in Tab. \ref{tab:text} and Fig. \ref{fig:text}.
Our model achieves the best FID score and user preference in the quantitative comparison.
The qualitative experiments show that our model can produce text-coherent results without ghost artifacts.
More cases can be seen in Fig. \ref{fig:figure-only-1}.

\begin{figure}[t]
    \centering
    \includegraphics[width=0.5\textwidth]{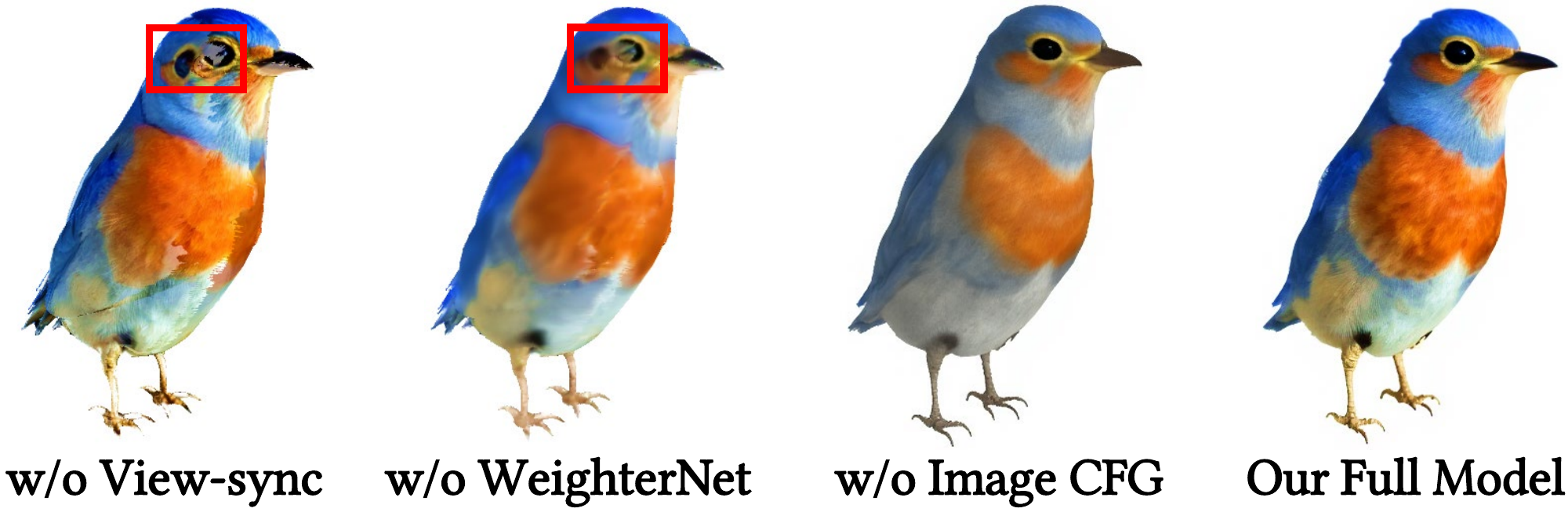}
    \caption{
        Qualitative results of our ablation study.
        Our full model achieves consistent and high-quality generation results, while the ablated methods suffer from ghosting artifacts and degraded quality.
    }
    \label{fig:sync}
\end{figure}

\begin{figure}[t]
    \centering
    \includegraphics[width=0.5\textwidth]{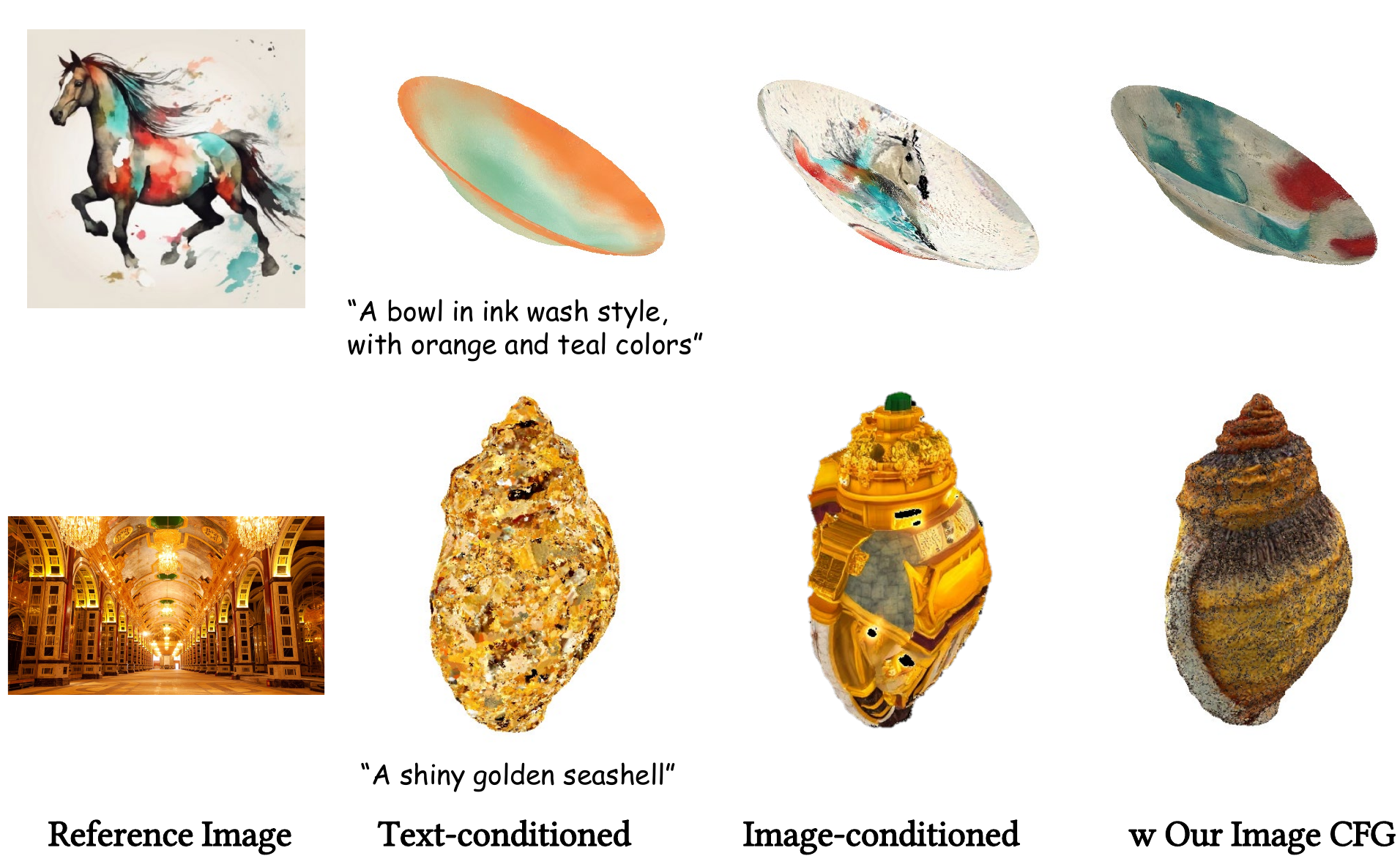}
    \caption{
        Comparison on the stylization strategy.
    }
    \label{fig:ab_style}
\end{figure}

\subsection{Comparisons of Image-to-Texture}

We conduct quantitative experiments on image-to-texture generation using the rendered images as input.
We use fixed trigger words as text prompts and set $\alpha_0=1$ for the image embedding. 
The results presented in Tab. \ref{tab:image} show that our method significantly outperforms others, indicating better image understanding and stronger generative capabilities.
For the qualitative experiments, we use imperfectly-aligned, in-the-wild images as conditions to test the generalizability of our method.
From the results shown in Fig. \ref{fig:image}, we can tell that our method can infer semantic correspondences from the image information to generate results that meet both image and geometry conditions.
More cases can be seen in Fig. \ref{fig:figure-only-2}.

\begin{figure*}[t]
    \centering
    \includegraphics[width=1\textwidth]{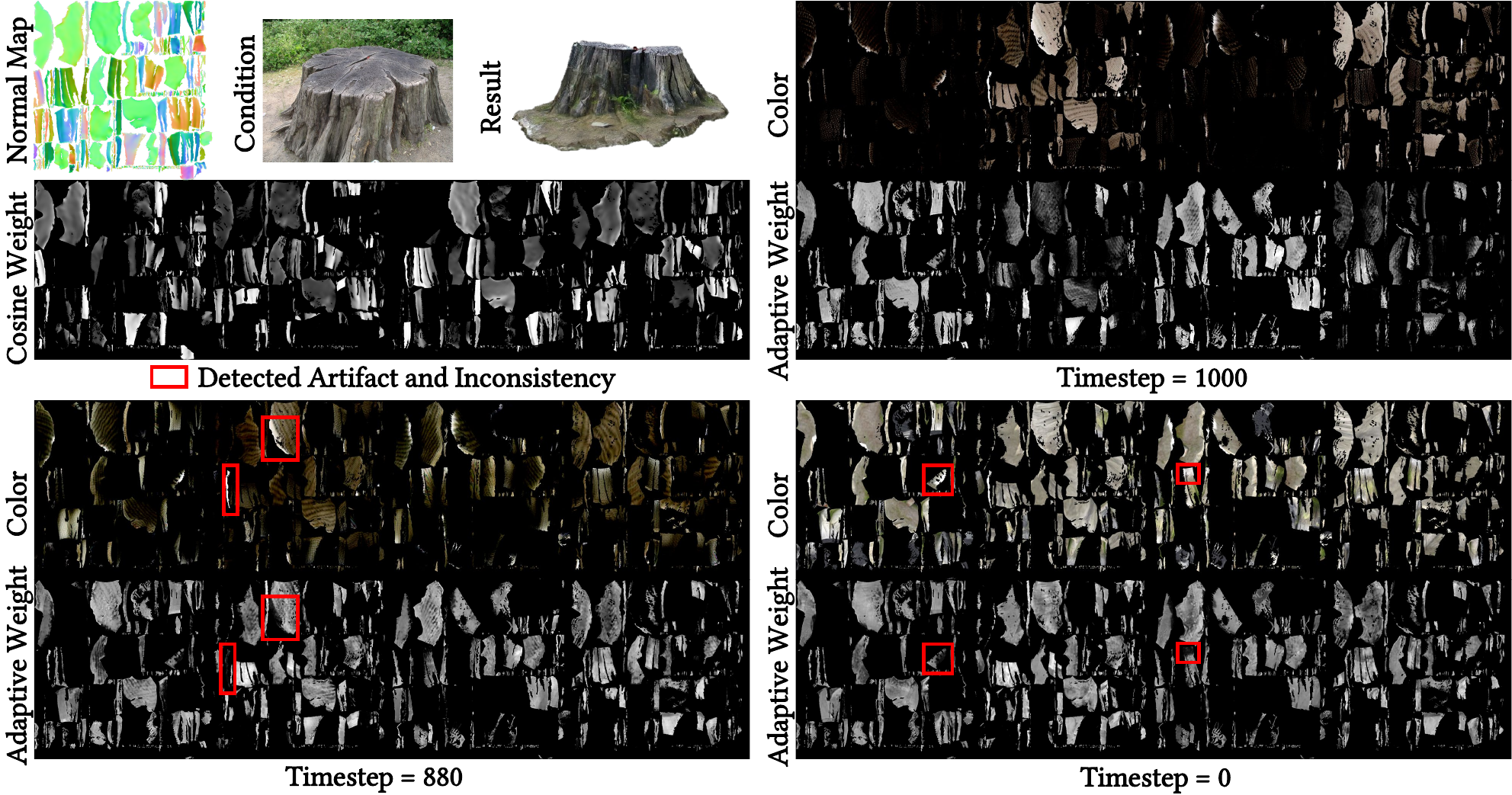}
    \caption{
        Analysis of the intermediate results from our WeighterNet.
        Weights and colors from different views are spliced horizontally.
        The first three pictures shows the normal, condition image and the render of the final generated results.
        Below shows the traditional cosine weights.
        We can see that it lacks generalizability and fails to capture misalignments, causing blurry output.
        Differently, images of the other three corners shows the color UV map and the corresponding adaptive weights of different diffusion time steps.
        From these images we can tell that the adaptive weights can detect misalignments and solve in a learning-based, generalizablr manner.
    }
    \label{fig:ab_wn}
\end{figure*}

\subsection{Ablation Studies}

We conduct ablation studies to demonstrate the effect of our view synchronization, WeighterNet (replaced by simple cosine weight), and the image-based CFG (replaced by normal text-based CFG) modules.
Quantitative and qualitative results can be seen in Table \ref{tab:ablation} and Fig. \ref{fig:sync}, respectively.
Although our multi-view generation model may still exhibit local misalignment in the UV space, our synchronization strategy and the 3D-aware weighting network can solve such a problem and generate flawless texture maps.
Due to multiple encode-decode operations during the sampling process, the original text-based CFG can lead to a decline in generation quality.
Meanwhile, our image-based CFG method can enhance condition consistency and generate higher-quality results.

We also conduct qualitative ablation studies on our WeighterNet and image-CFG-based stylization. The results are visualized in Fig. \ref{fig:ab_style} and Fig. \ref{fig:ab_wn}.
Compared with directly using text descriptions for stylized texture generation, using reference images can improve the accuracy of style control.
However, the structure and semantic information implied in the image may conflict with the mesh, resulting in the generated texture containing the semantic information in the original image.
The comparison is in Fig. \ref{fig:ab_style}.
Fig. \ref{fig:ab_wn} shows that our WeighterNet has the ability to assign weight adaptively to different views at different sample timestep, while the traditional view angle-based method only calculates fixed weights and fails to detect inconsistency and bad generation.

\begin{figure*}[t]
    \centering
    \includegraphics[width=1\textwidth]{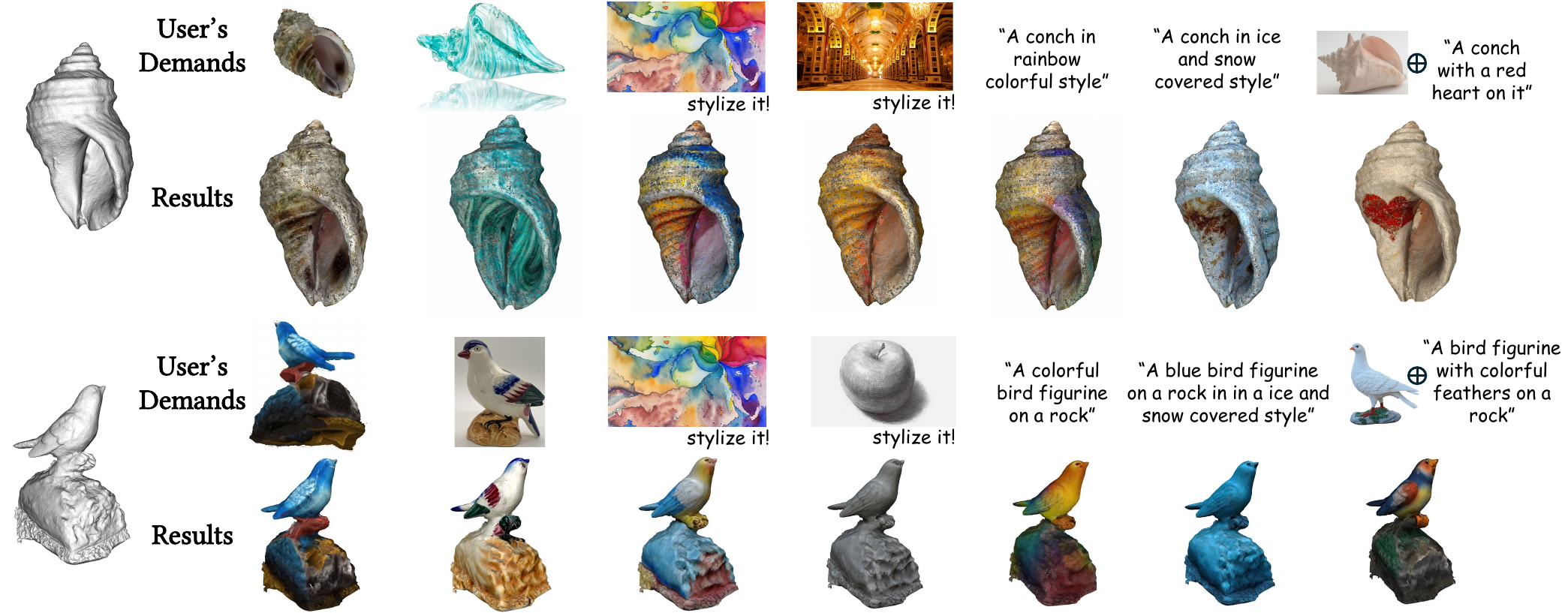}
    \caption{
        Our Applications includes tasks of text-to-texture, image-to-texture, text-guided image refinement, and reference image-based stylization.
        More cases of our applications including the cases shown in the teaser can be seen in the appendix.
    }
    \label{fig:app}
\end{figure*}

\subsection{Applications}

Our method can achieve various applications, including tasks of text-to-texture, image-to-texture, multi-modal joint generation, and stylization using a reference image.
The results are shown in Fig. \ref{fig:inter} and Fig. \ref{fig:app}.
These results show that our method can cope with different user needs with different prompts in actual production and flexibly generate high-quality textures that meet user intentions.
Specifically, for the multi-modal joint generation, we have two different modes.
If the text and image prompt provided by the users are describing the same thing, the joint generated texture will be a mixture of the two modality (as shown in Fig. \ref{fig:inter}).
Alternatively, if the text serves as a detailed supplement to the image, text-guided image refinement can be achieved. (as shown in the rightmost column of Fig. \ref{fig:app}).

\begin{table}[t]
\centering
\caption{
    Quantitative ablation studies on our view synchronization, WeighterNet, and image-based CFG.
    The best results are indicated in bold.
    KID multiplied by ${10}^{-4}$.
}

\resizebox{1.0\linewidth}{!}{
\begin{tabular}{ccccc}
\hline
& \makecell[c]{w/o\\view-sync} 
& \makecell[c]{w/o\\WeighterNet} 
& \makecell[c]{w/o\\image CFG} 
& \makecell[c]{our\\full model} \\
\hline
FID $\downarrow$ & 80.635 & 76.688 & 79.041 & \textbf{71.621} \\
KID $\downarrow$ & 77.883 & 59.674 & 65.482 & \textbf{58.465} \\
\hline
\end{tabular}
}
\label{tab:ablation}
\end{table}

\section{Conclusion}

We introduce FlexPainter, a novel framework for generating high-quality, consistent texture maps with flexible generation.
We achieve multi-modality mixing and flexible conditional control by constructing a shared semantic space.
The proposed image-based CFG also enables structure suppression of the reference image and achieves high-fidelity stylization.
For texture coherency, we use a multi-view image grid as a generation target for global understanding. 
The reprojection-based view synchronization and weighting module reinforces multi-view alignment for local consistency. Finally, a texture completion and enhancement module produces high-resolution textures. Extensive experiments show that FlexPainter outperforms existing methods, offering superior semantic alignment and multi-view consistency, making it a practical solution for 3D artists and designers.
Finally, a texture completion and enhancement module generates full high-resolution textures, significantly improving the visual quality and detail of the final output.

{
    \small
    \bibliographystyle{unsrt}
    \bibliography{main}
}

\twocolumn[{
\renewcommand\twocolumn[1][]{#1}
\begin{center}
    {\fontsize{14pt}{18pt}\selectfont \bfseries Supplementary Material for \\
    FlexPainter: Flexible and Multi-View Consistent Texture Generation \par}
\end{center}
}]
\setcounter{section}{0}
\setcounter{figure}{0}
\setcounter{table}{0}
\renewcommand\thesection{S\arabic{section}}
\renewcommand\thefigure{S\arabic{figure}}
\renewcommand\thetable{S\arabic{table}}

\appendix

\section{Additional Implementation Details}

\subsection{Baking Texture}

When preparing for texture data, we found one problem in Objaverse\cite{deitke2023objaverse}.
Instead of a single unified texture, some meshes are composed of multiple parts, each with its texture image.
For those meshes, we use Blender's Smart UV Project function to re-unfold the UVs and then bake the diffuse color to get a unified texture image.
For the texture completion net, we bake 10K texture maps at 1K ($1024^2$ pixels) resolution; for the texture enhancement net, we bake the same texture maps at a 4K ($4096^2$ pixels) resolution.

\subsection{Training of LoRA.}

During LoRA training, we use Adam \cite{kingma2014adam} optimizer with a learning rate of $10^{-5}$.
The LoRA rank is set to 256.
We train the model for 20K steps with 8 NVIDIA-A800 GPUs and a batch size of 4.
We use "'a grid of 2x2 multi-view image. white background." for the trigger word.

\subsection{Training of WeighterNet.}

For the training of WeighterNet, we use Adam optimizer with a learning rate of $10^{-4}$.
The weight of the supervision loss are set to: $\lambda_{pec} = 1, \lambda_{cyc} = 0.5$.
Additionally, a smooth loss $L_{sm}$ with weight $\lambda_{sm} = 0.2$ is used to regularize the training process:
\begin{equation}
    L_{sm} = \sum ((\nabla_x I)_{i, j}^2+\left(\nabla_y I\right)_{i, j}^2),
\end{equation}
and the full supervision loss $L$ is set to:
\begin{equation}
    L = \lambda_{perc} L_{perc} + \lambda_{cyc} L_{cyc} + \lambda_{sm} L_{sm} 
\end{equation}
We train the model for 100K step with 8 NVIDIA-A800 GPU and batchsize of 2.

\subsection{Training of Completion and Enhancement Net.}

For texture completion net, we start from the checkpoint provided by TEXGen \cite{yu2024texgen}.
However, since the original TEXGen is trained using a partial texture map reprojected from a single view, direct extension into a larger partial UV map from four views may result in domain variance.
To this end, we generate a partial texture map using our four-view camera setting and finetune the original model.
We use an Adam optimizer with a learning rate of $10^{-4}$ and train the model for 20K steps on 8 NVIDIA-A800 GPUs with a batch size 2.

For texture enhancement net, we choose Real-ESRGAN\cite{wang2021real} as our base model.
Given 4K (4096 × 4096) baked texture maps, we generate corresponding low-quality images with the degradation process described in Real-ESRGAN\cite{wang2021real}.
During training, we use Adam \cite{kingma2014adam} optimizer with learning rate of $10^{-4}$. 
Additionally, we utilize L1 Loss, GAN Loss, and Perceptual Loss for pixel level, content level, and style level supervision:
\begin{equation}
    L = \lambda_{perc} L_{perc} + \lambda_{p} L_{p} + \lambda_{g} L_{g} 
\end{equation}
where $\lambda_{perc}, \lambda_{p}, \lambda_{g}$ are set to 1.0, 1.0, 0.1 respectively.
We train the model for 400K step with 1 NVIDIA-A800 GPU and batchsize of 12.

\begin{figure*}[!t]
    \centering
    \includegraphics[width=0.87\textwidth]{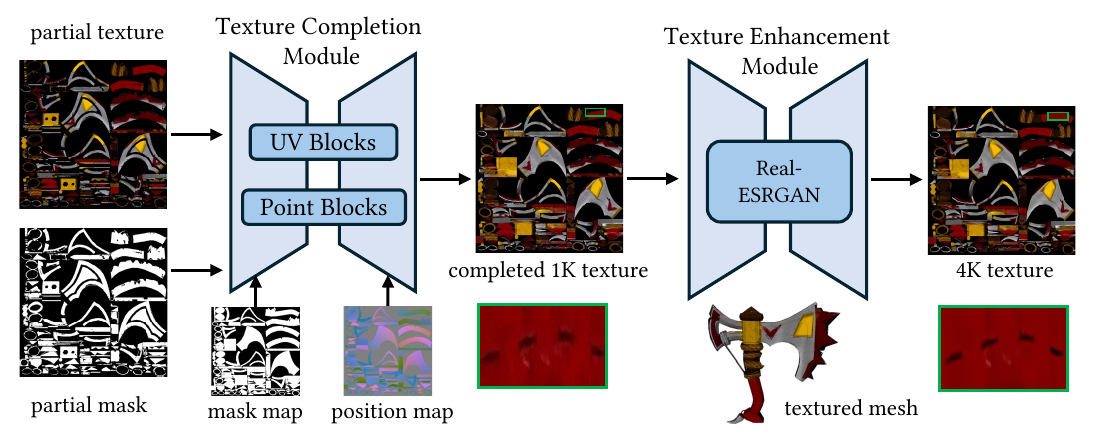}
    \caption{
        Detailed pipeline of the texture completion and enhancement module.
        We use a 3D-aware texture completion module containing UV blocks and point blocks to complete the partial texture.
        Then, we use a texture enhancement module to generate the super-resolution results.
    }
    \label{fig:stage2}
\end{figure*}

Detailed pipeline of the texture completion and enhancement module is shown in Fig.\ref{fig:stage2}.

\subsection{Implementaion of CFG}

Since the original checkpoint from FLUX.1-dev \cite{flux2023} uses a distilled CFG mechanism, we implement an explicit CFG method following \cite{ho2022classifier} during sampling.
For all of our experiments, the distilled CFG scale is set to 6, and the explicit CFG scale is set to 2.
We use the embedding of a white image as the negative embedding for the explicit CFG.

\section{More Qualitative Results}
\begin{figure*}[t]
    \centering
    \includegraphics[width=0.95\textwidth]{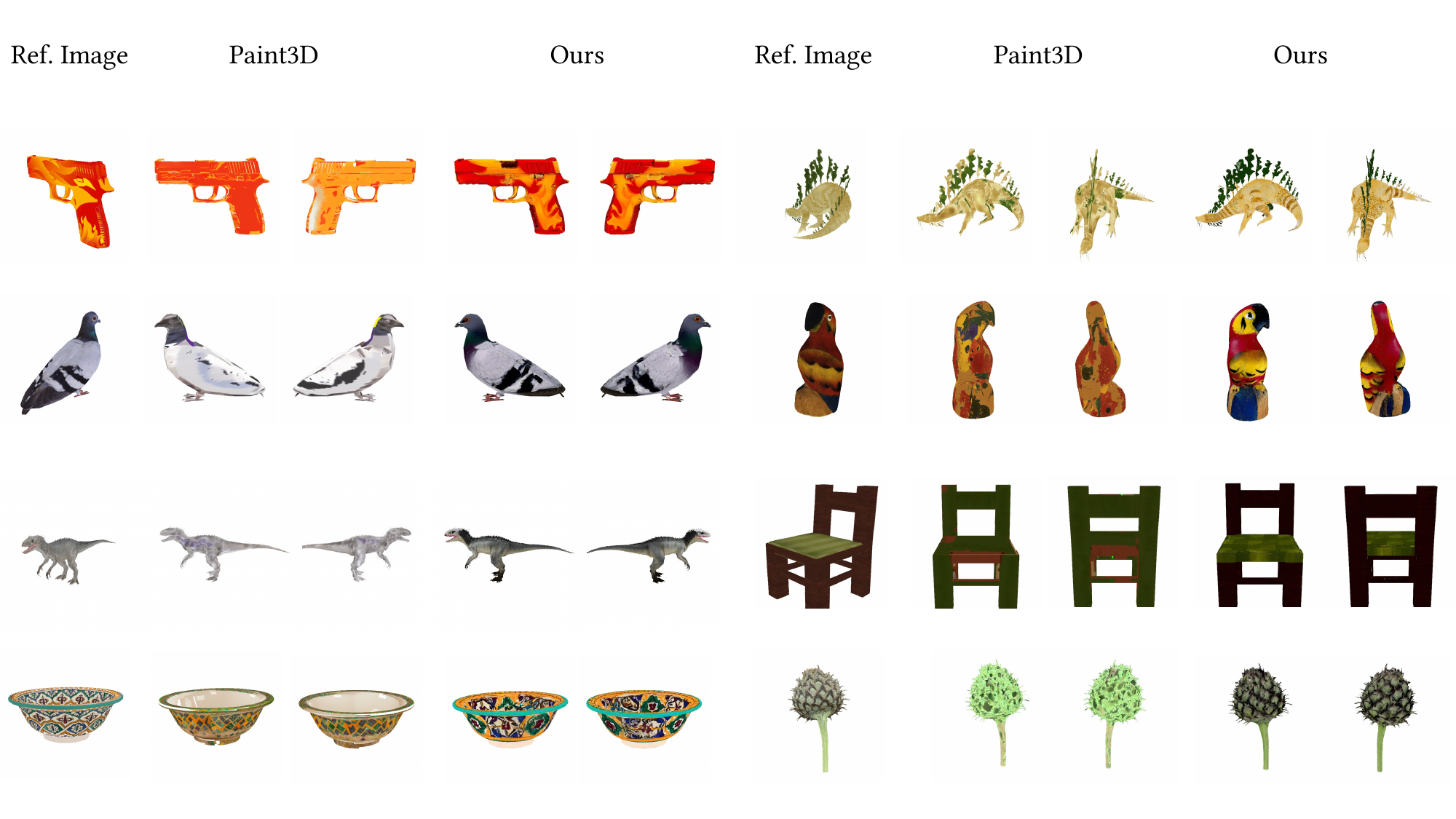}
    \caption{
        More cases of our image-to-texture generation.
    }
    \label{fig:supp_image}
\end{figure*}

\begin{figure*}[t]
    \centering
    \includegraphics[width=1.0\textwidth]{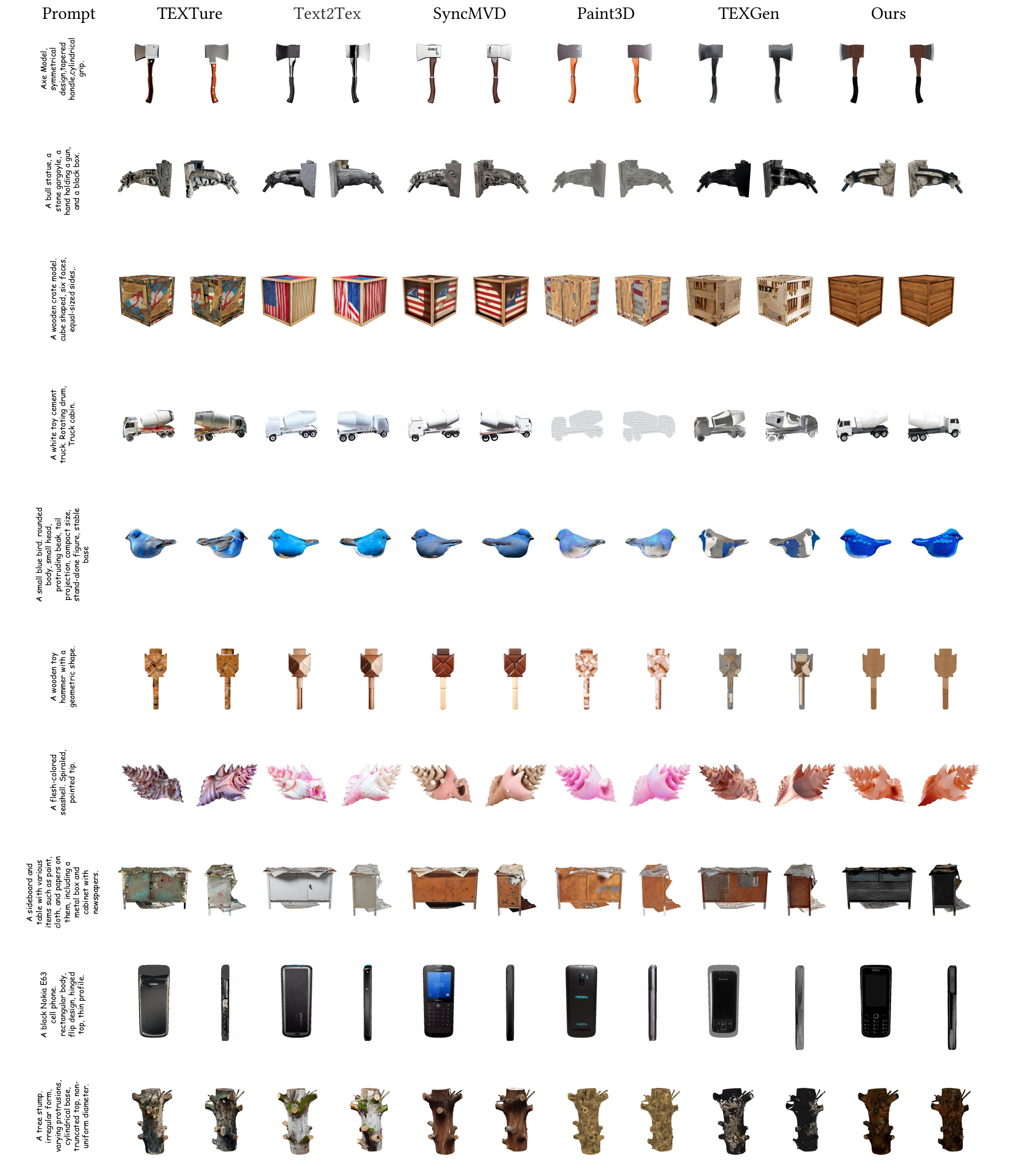}
    \caption{
        More cases of our text-to-texture generation.
    }
    \label{fig:supp_text}
\end{figure*}

\begin{figure*}[t]
    \centering
    \includegraphics[width=0.95\textwidth]{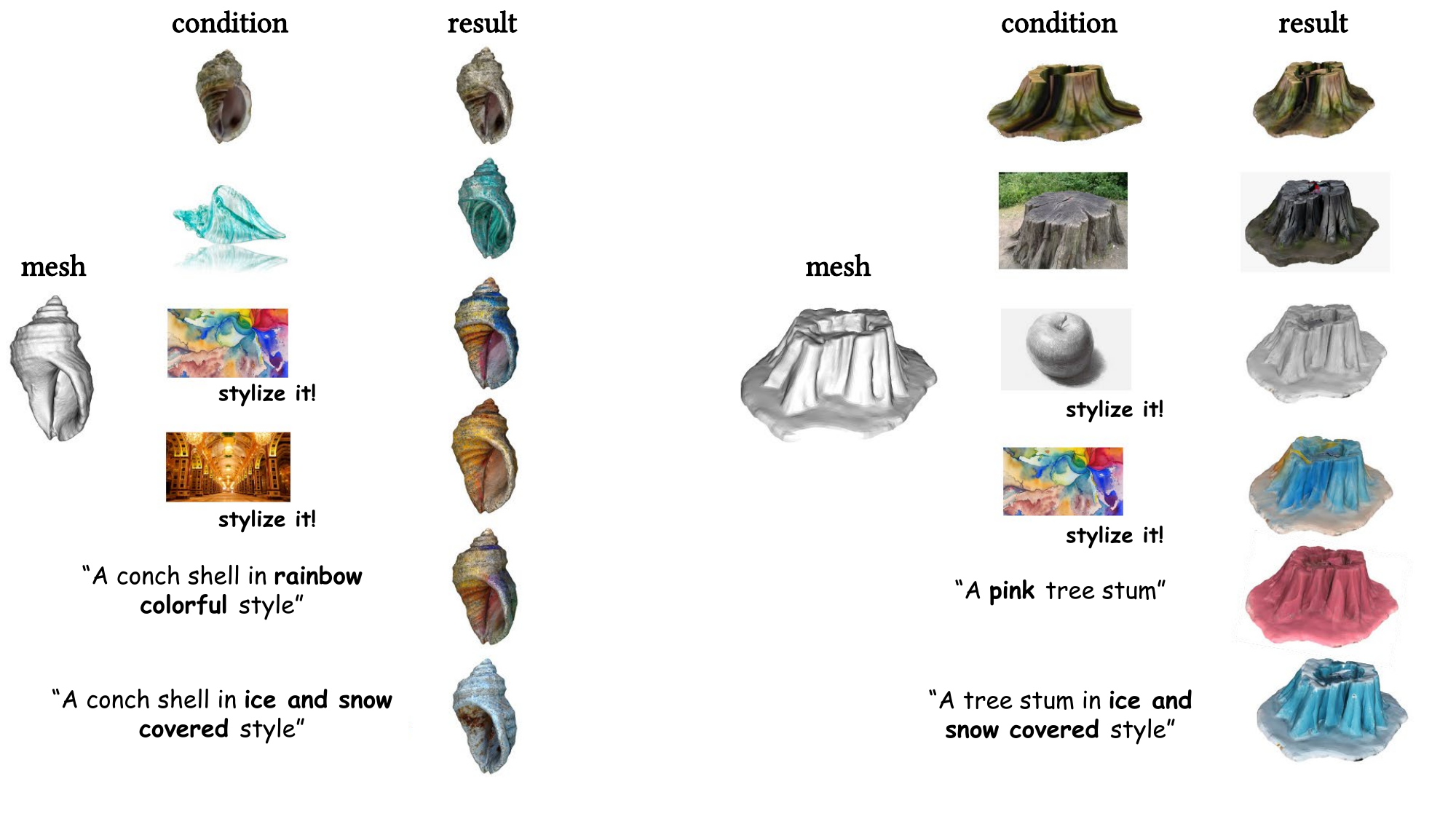}
    \caption{
        More cases of our applications.
    }
    \label{fig:supp_teaser_start}
\end{figure*}

\begin{figure*}[t]
    \centering
    \includegraphics[width=0.95\textwidth]{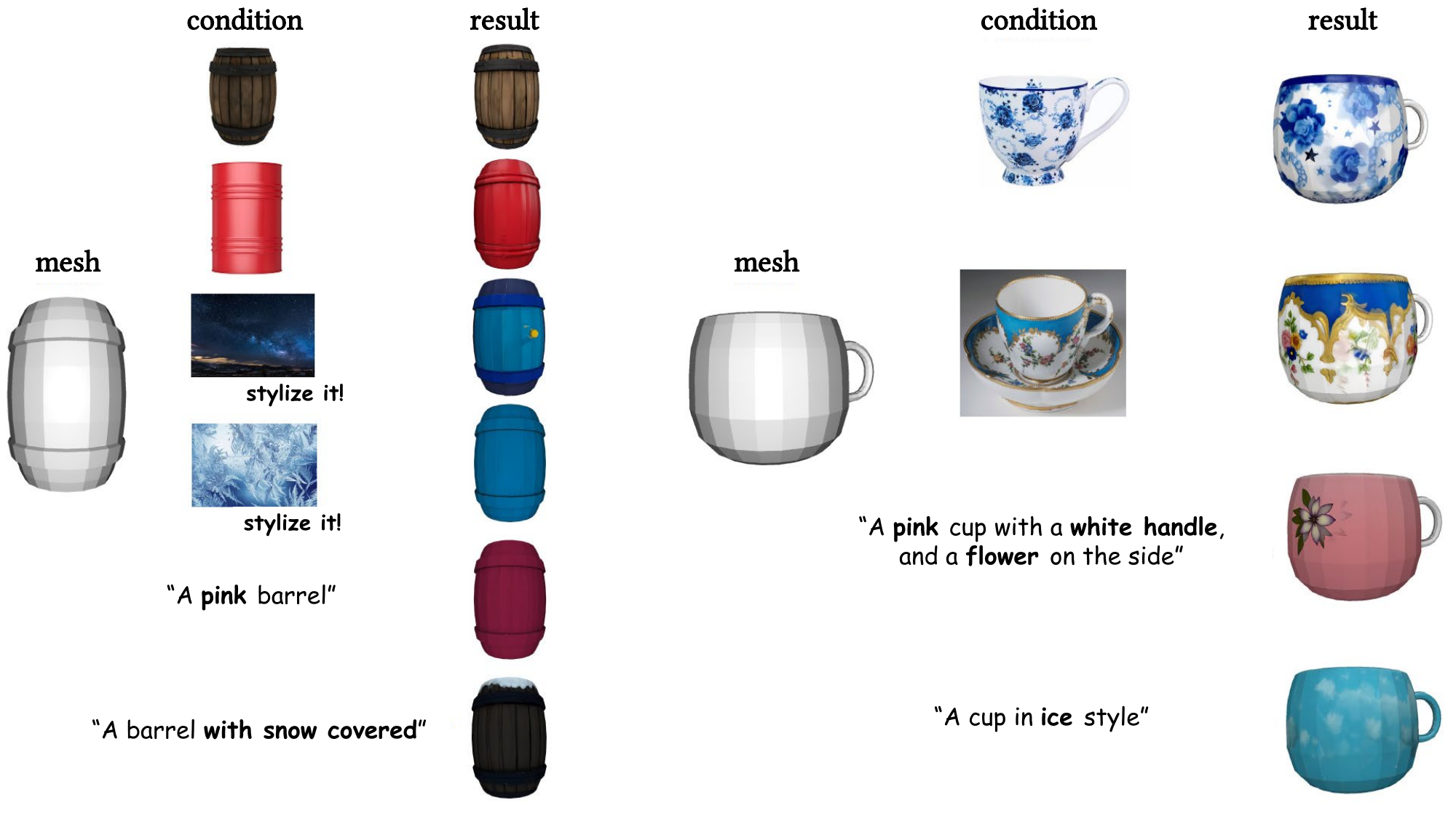}
    \caption{
        More cases of our applications.
    }
    \label{fig:supp_teaser_end}
\end{figure*}

Our main paper introduces various applications with our model, including text-to-texture, image-to-texture, and texture stylization. 
We demonstrate more results in Fig.\ref{fig:supp_text}, Fig.\ref{fig:supp_image}.
Additionally in Fig.\ref{fig:supp_teaser_start} to Fig. Fig.\ref{fig:supp_teaser_end}, we detailed conditions and generation results shown in the teaser of the main paper.
Also, we attached a video to the supplementary file to present the texture results using a dynamic approach.

\section{Details of our user study}

In our user study, we use 120 cases for text-to-texture generation and 50 cases for image-to-texture generation.
We shuffle all the cases and separate them randomly into 6 questionnaires and distribute them to 100 interviewees.

\section{Limitations and Future Work}

Although our model has already achieved superior results, some challenges remain to be solved.
While utilizing depth maps as geometric conditions is effective, it often results in the loss of geometric detail from the original mesh.
This discrepancy can lead to the generation results in Stage I not fully aligning with the original mesh.
Moreover, we are still exploring whether explicit 3D information can be used as guidance to steer the inpainting process in Stage II.
Due to the albedo data used during training, our generated textures are free from ambient lighting.
However, we have observed that applying these textures in practical 3D generation pipelines still requires designers to spend considerable time adjusting the lighting. Therefore, our future work will explore methods to generate textures that align with designers' expected lighting effects to achieve more flexible, thereby reducing design time.

\subsection{Texture Completion and Enhancement}

We generate and blend the multi-view images in the previous stage, resulting in a single UV map $\tilde{T}^0$.
Although most of the object surface is covered, some regions still require texture completion due to occlusion.
A straightforward approach is to use an image diffusion model to perform texture inpainting in the UV space.
However, due to the discontinuity inherent in UV space, this method lacks an understanding of the spatial relationships, potentially leading to suboptimal results.
Leveraging the 3D-aware architecture of TEXGen \cite{yu2024texgen}, we finetune it with our aggregated four-view UV map.
Due to the explicitly introduced 3D knowledge and the texel-aligned input, our texture completion network produces reasonable and spatially correlated inpainting in the UV space.

Although our method can generate high-quality, 1K ($1024^2$ pixels) resolution texture maps through the above pipeline, some scenarios require a higher resolution for more detailed rendering.
To address this limitation, we perform super-resolution on the generated texture maps.
We finetune a Real-ESRGAN \cite{wang2021real} model using our baked texture maps to perform 4 times up-scale.
At can be produced.
At last, precise and detailed texture maps at a resolution of 4K ($4096^2$ pixels).

\begin{figure*}[!t]
    \centering
    \includegraphics[width=0.98\textwidth]{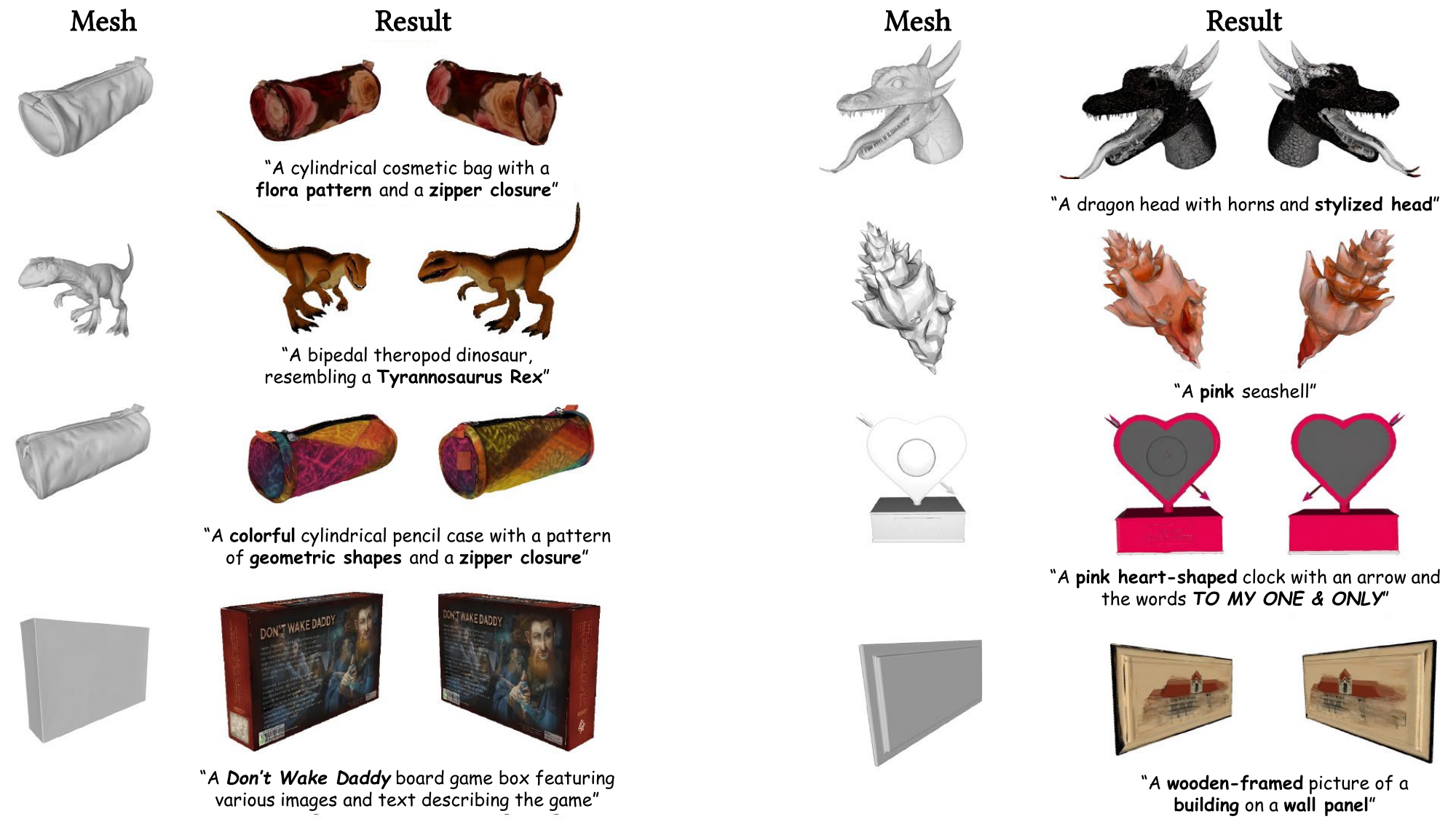}
    \caption{
        More cases of our text-to-texture generation.
    }
    \label{fig:figure-only-1}
\end{figure*}

\begin{figure*}[!t]
    \centering
    \includegraphics[width=0.98\textwidth]{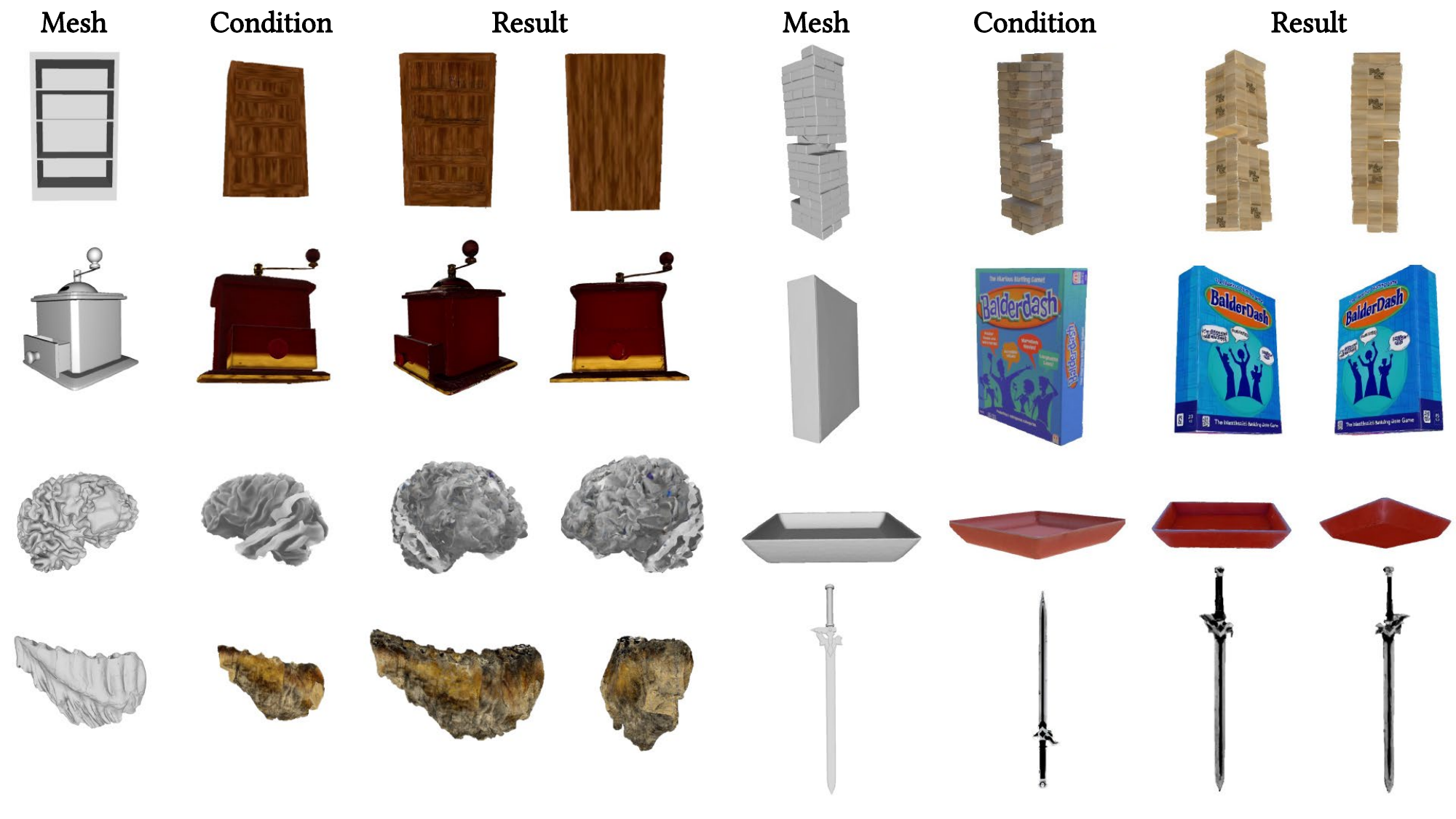}
    \caption{
        More cases of our imgae-to-texture generation.
    }
    \label{fig:figure-only-2}
\end{figure*}

\begin{figure*}[t]
    \centering
    \includegraphics[width=0.98\textwidth]{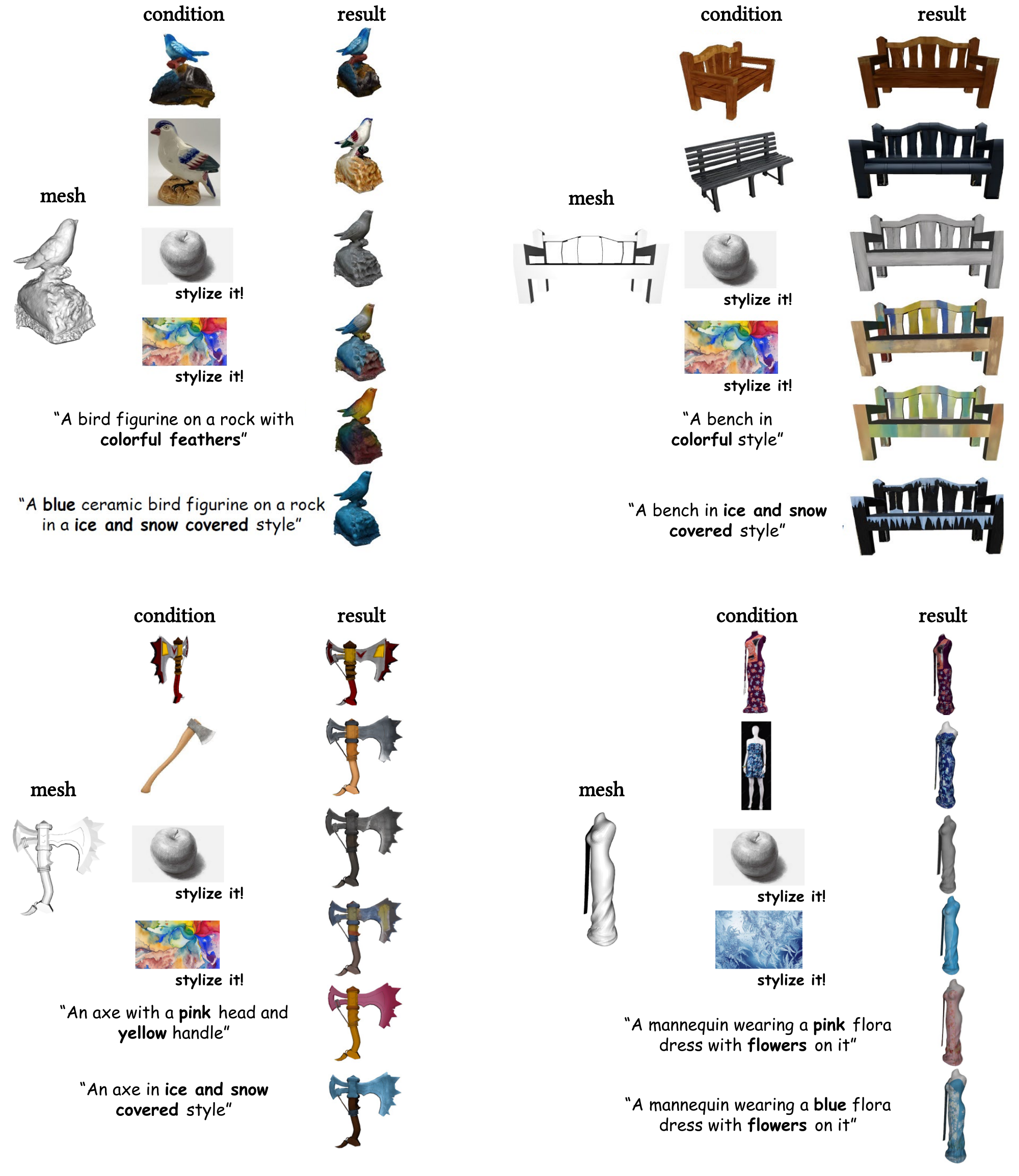}
    \caption{
        More cases of our applications.
    }
    \label{fig:figure-only-3}
\end{figure*}

\begin{figure*}[t]
    \centering
    \includegraphics[width=0.95\textwidth]{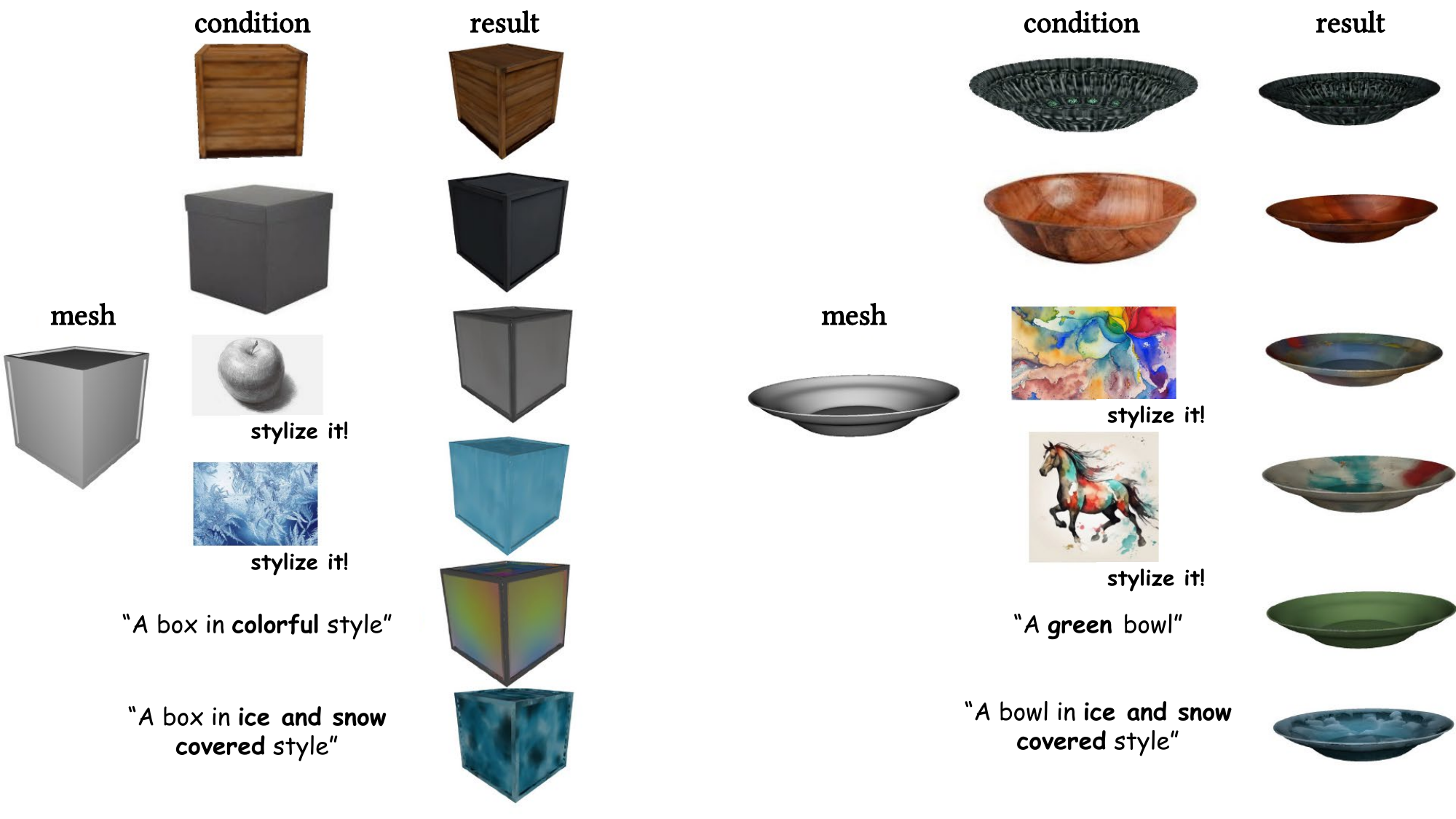}
    \caption{
        More cases of our applications.
    }
\end{figure*}

\end{document}